\documentclass[graybox]{svmult}

\usepackage{mathptmx}       

\usepackage{helvet}         
\usepackage{courier}        

\usepackage{graphicx}              

\usepackage{siunitx}
\usepackage{amsbsy}
\usepackage{amsmath,bm}
\usepackage{amssymb}
\usepackage{amsfonts}

\usepackage{multirow}

\usepackage[bottom]{footmisc}
\usepackage{cite}

\usepackage{url}

\begin{document}

\title*{Evaluation of Torque Ripple and Tooth Forces of a Skewed PMSM by 2D and 3D FE Simulations}
\titlerunning{Evaluation of Torque Ripple and Tooth Forces}
\author{Karsten Müller\inst{1}
	\and
	Andreas Wanke\inst{1}
	\and
	Yves Burkhardt\inst{2}
	\and
	Herbert De Gersem\inst{2}}
	
\authorrunning{Evaluation of Torque Ripple and Tooth Forces of a Skewed PMSM by 2D and 3D FEA}

\institute{Mercedes-Benz AG2, 70327 Stuttgart, Germany \
	\texttt{karsten.k.mueller@mercedes-benz.com}, 
	\texttt{andreas.wanke@mercedes-benz.com}
	\and Department of Electrical Engineering and Information Technology, Technical University Darmstadt, 64283 Darmstadt, Germany \
	\texttt{yves.burkhardt@eas.tu-darmstadt.de},
	\texttt{degersem@temf.tu-darmstadt.de}
.}
\maketitle
\abstract*{In this paper, various skewing configurations for a permanent magnet synchronous machine are evaluated by comparing torque ripple amplitudes and tooth forces. Since high-frequency pure tones emitted by an electrical machine significantly impact a vehicle's noise, vibration, and harshness (NVH) behavior, it is crucial to analyze radial forces. These forces are examined and compared across different skewing configurations and angles using the Maxwell stress tensor in 2D and 3D finite-element (FE) simulations.
	In addition to conventional investigations in 2D FE simulations, 3D FE simulations are executed. These 3D FE simulations show that axial forces occur at the transition points between the magnetic segments of a linear step skewed rotor.}

\abstract{In this paper, various skewing configurations for a permanent magnet synchronous machine are evaluated by comparing torque ripple amplitudes and tooth forces. Since high-frequency pure tones emitted by an electrical machine significantly impact a vehicle's noise, vibration, and harshness (NVH) behavior, it is crucial to analyze radial forces. These forces are examined and compared across different skewing configurations and angles using the Maxwell stress tensor in 2D and 3D finite-element (FE) simulations.
	In addition to conventional investigations in 2D FE simulations, 3D FE simulations are executed. These 3D FE simulations show that axial forces occur at the transition points between the magnetic segments of a linear step skewed rotor.}

\section{General Instructions}
\label{sec:general}
For automotive applications, permanent magnet synchronous motors (PMSMs) are widely used due to their advantages, such as high efficiency and high power density \cite{Wang.2020}. PMSMs must also meet the noise, vibration, and harshness (NVH) requirements essential for driving comfort \cite{Wang.2020}. NVH issues are primarily caused by torque ripples and tooth forces, which can lead to deformations of the stator yoke and housing as well as bearing fatigue \cite{Wang.2020}. The main contributors are radial magnetic forces acting on the stator, which follow the spatial and temporal distribution of the magnetic flux density in the air gap. A commonly employed and effective mitigation technique to address these issues is skewing the rotor or the stator slots. 
The beneficial effects of skewing on reducing cogging torque have been extensively analyzed in previous studies, which predominantly rely on 2D finite-element (FE) simulations to characterize the reduction in torque ripple amplitudes due to skewing \cite{Pile.2018b}, \cite{Park.2016b}, \cite{Paul.2019}. This paper compares the results of 2D and 3D FE simulations, focusing on torque harmonics and tooth forces for different skewing configurations. Special attention is given to the local force distribution on the stator. Furthermore, the influence of axial flux components on tooth forces is quantified using 3D FE simulations, highlighting the added value and accuracy of 3D FE approaches.

\section{Rotor skewing in a PMSM}
\begin{figure}
	\begin{center}
		
		\includegraphics[width=0.9\textwidth]{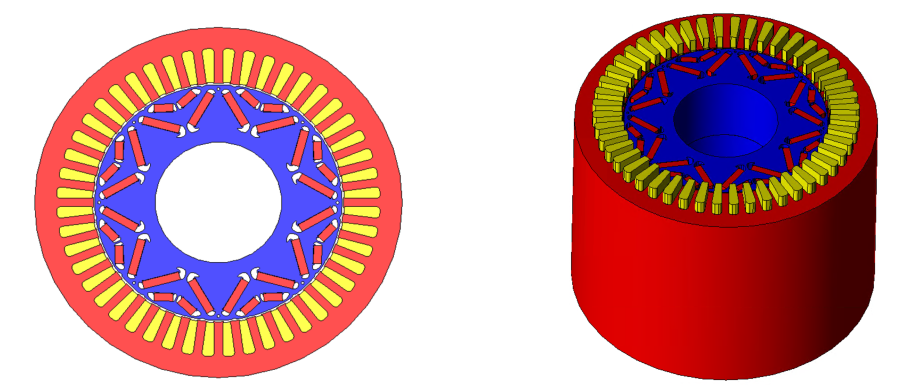}
	\end{center} 
	\caption{Visualisation of the RFM in 2D FE simulations (left) and 3D FE simulations (right)}
	\label{fig:RFM-2D3D} 
\end{figure}

The motor under investigation is an 8-pole, 48-slot PMSM with a double V-arrangement of magnets, as shown in Fig.~\ref{fig:RFM-2D3D}. For 2D FE simulations, the skewing of individual segments can be accounted for by either shifting the starting position of the rotor or adjusting the pole angle during post-processing. This approach allows the simulation of a single segment, significantly reducing computational effort. In this study, however, all segments were simulated individually in 2D and their results were superimposed during post-processing. 
For the 3D FE simulations, each segment was explicitly modeled, including simplified winding heads to account for end effects through an axial extension of the windings beyond the edge of the stator. The 2D FE simulations used approximately 12000 mesh elements per segment, whereas the full 3D FE model, which includes the complete axial length, consisted of approximately 1.5 million mesh elements.

\begin{figure}
	\begin{center}
		
		\includegraphics[width=0.9\textwidth]{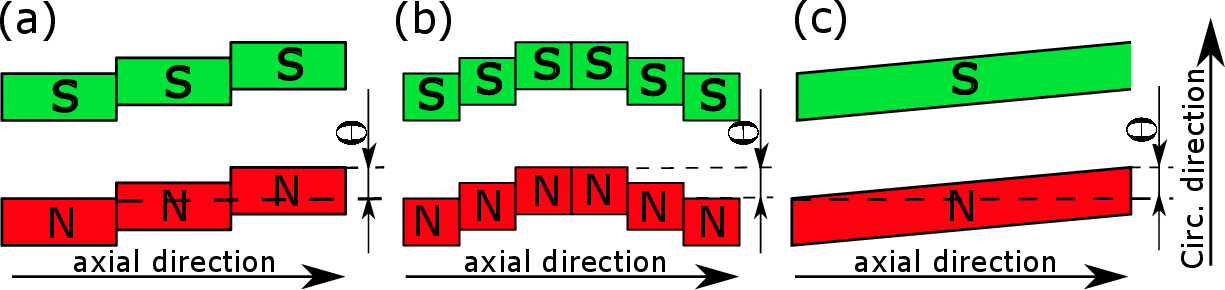}
	\end{center} 
	\caption{Diagram of different rotor-skewing configurations \\ (a) Step Skewing, (b) V-Skewing, (c) Continuous Skewing}
	\label{fig:Rotor-skewing-structures} 
\end{figure}

Figure~\ref{fig:Rotor-skewing-structures} illustrates different skewing configurations for a RFM. In step skewing, $q$ segmented permanent magnets are arranged in axial direction, each shifted by a fraction $\frac{\theta}{q}$ of the total skewing angle $\theta$. In V-skewing, the principle is applied in a symmetric way. In continuous skewing, a single permanent magnet is mounted with a circumferential inclination angle $y = \frac{r_{rt}\theta}{L}$ with $L$ is the rotor length and $r_{rt}$ the rotor radius. Continuous skewing is the limit case of step skewing where $q$ $\rightarrow$ $\infty$ \cite{Binder.2017b}.
The optimum skewing angle $\theta_{\mathrm{skew}}$ is calculated from the number of poles $N_p$ and the number of stator slots $N_s$ by
\begin{equation}
	\theta_{\mathrm{skew}} = \frac{2\pi}{N_c}
\end{equation}
where $N_c$ is the least common multiple of $N_p$ and $N_s$. 
Skewing the stator or the rotor by $\theta_{\mathrm{skew}}$ reduces the magnetic coupling between stator and rotor, and thus the flux linkage and the induced voltage, by the skew factors \cite{Binder.2017b}
\begin{equation}
	\kappa_{\mathrm{skew,\nu}} = \frac{\sin \left( \frac{\theta_{\mathrm{skew}} \nu}{2} \right) }{\frac{\theta_{\mathrm{skew}} \nu}{2}},
\end{equation}
which depend on the harmonic orders
\begin{equation}
	\nu = 1 \pm g \frac{N_s}{N_p}, \space \newline
	g \in \mathbb{N}.
\end{equation}
According to equation (1), the considered PMSM with 48 slots and 8 poles has a $N_c$ of $48$ and therefore a total skewing angle of $7.5^\circ$.

\section{Analysis of the torque ripples considering rotor skewing}
\begin{figure}[b]
	\centering
	\includegraphics[width=0.9\textwidth]{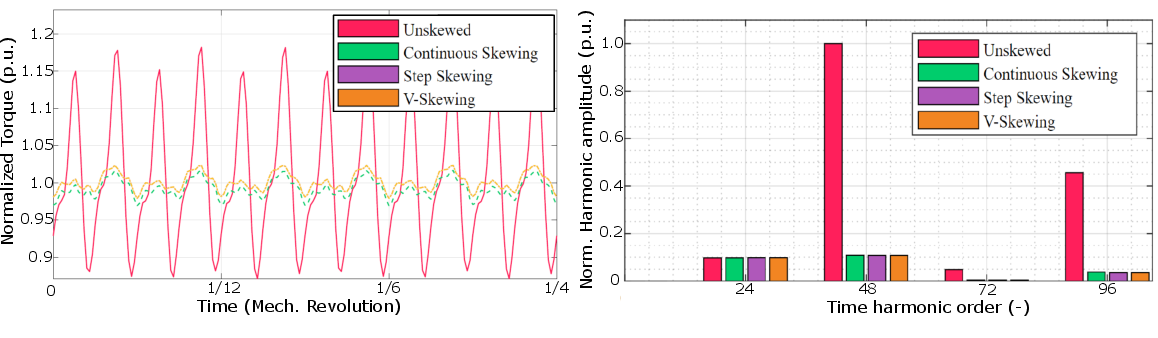}
	\caption{Torque over mechanical rotor angle for different rotor skewing configurations at load condition (left) and corresponding Fourier transform of the torque (right)}
	\label{fig:TorqueRipples-load} 
\end{figure}

\begin{figure}
	\begin{center}
		
		\includegraphics[width=0.75\textwidth]{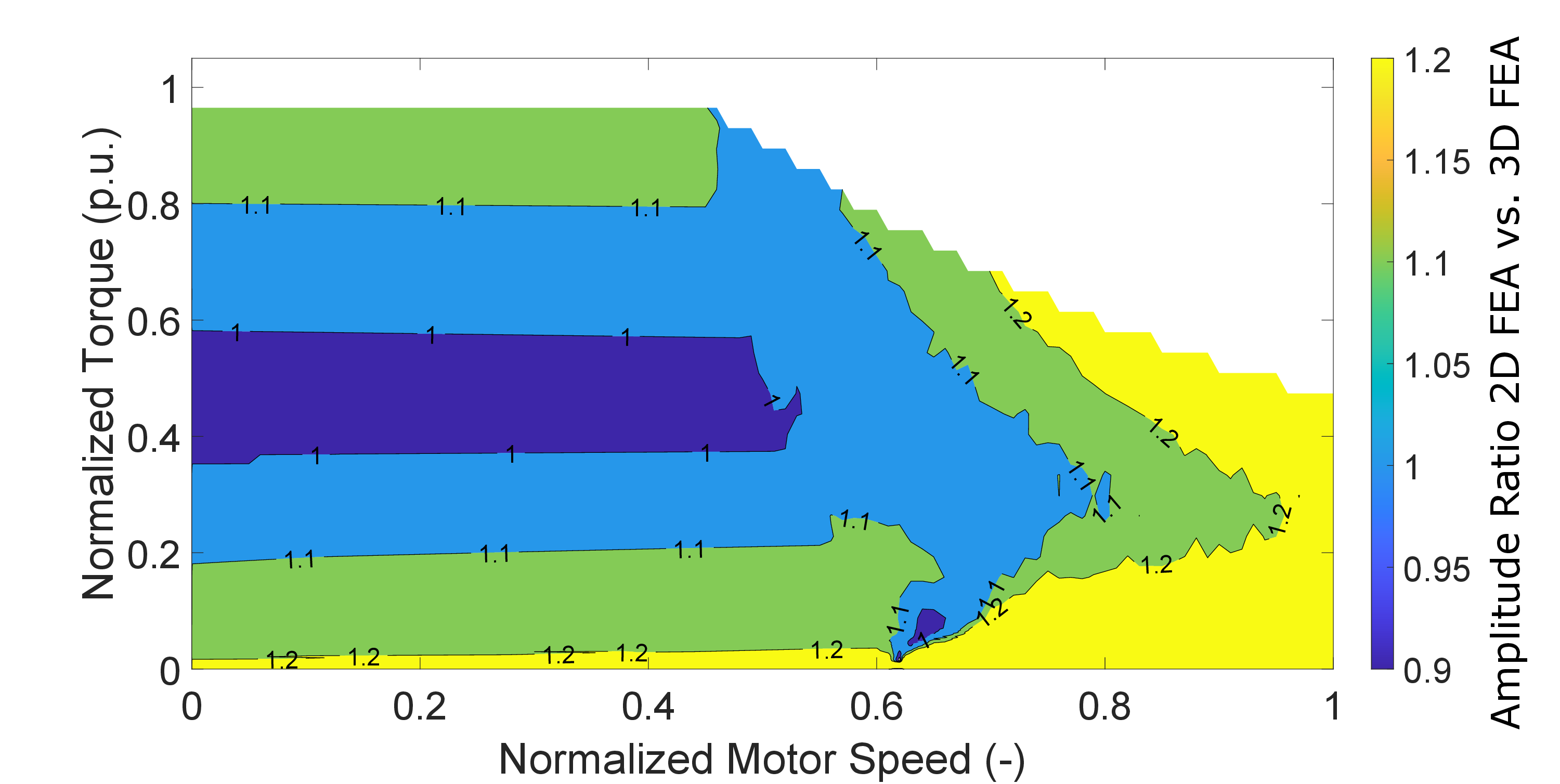}
	\end{center} 
	\caption{Ratio between 2D and 3D simulation results for the magnitude of the 48th torque harmonic, shown for all operation points in the motor map}
	\label{fig:TorqueRipples-Kennfeld} 
\end{figure}

Torque ripples in a PMSM result from the interaction of multiple harmonic components of the magnetic field. The total torque, including the ripple components, can be derived from the magnetic flux density in the air gap using the Maxwell stress tensor \cite{Yamazaki.2022}. The magnetic flux density $B$ in the air gap of rotating machines includes both time and space harmonic components, which can be expressed as a 2D Fourier series \cite{Harmonic,Binder.2017b}. 
Assuming a 2D symmetry, the torque derived using the Maxwell stress tensor is given by
\begin{equation}
	M_z = L \int_{0}^{2\pi}\frac{B_rB_\theta}{\mu_0} r_\mathrm{\delta}^2 \partial\theta
\end{equation}
with the axial length $L$ and $\theta$ the rotation angle in the air gap. In addition to the constant contributions to the torque, oscillating components arise from the combinations of unequal spatial and temporal harmonics. Since these necessarily include at least one order $\nu$ larger than 3, they appropriately decay in magnitude by the according skew factors. \\
Figure~\ref{fig:TorqueRipples-load} shows on the left side the torque curve for one electrical period obtained by 2D FE analysis, and on the right side the corresponding torque ripple amplitudes for the high-current operating point in the base speed range. Due to the different skewing configurations, the peak-to-peak amplitude of the torque decreases, which is also evident from the subsequent Fourier analysis.
Compared to previous literature, Fig.~\ref{fig:TorqueRipples-Kennfeld} shows the ratio between 2D and 3D simulations results for the 48th torque harmonic in the torque-speed characteristics diagram for a skewed rotor. For lower speed, the ratio is negligible. However, at higher speeds, it becomes significant. Similar findings are observed for other skewing configurations. One reason for the increasing deviation between the 2D and 3D FE simulation results with increasing motor speed is the presence of local effects and saturations that cannot be fully modeled in 2D. Additionally, stronger field weakening is required in 3D, which results in less q-current and thus a lower torque amplitude.  

As a conclusion, 2D FE simulation results for the torque are sufficiently accurate, except for the larger speed range. 
For calculation of the torque harmonics, 2D FE simulation results have generally a good alignment with 3D but for some operation modes 3D FE analysis is unavoidable.

\section{Analysis of tooth forces considering the rotor skewing}
\begin{figure}
	\begin{center}
		
		\includegraphics[width=0.5\textwidth]{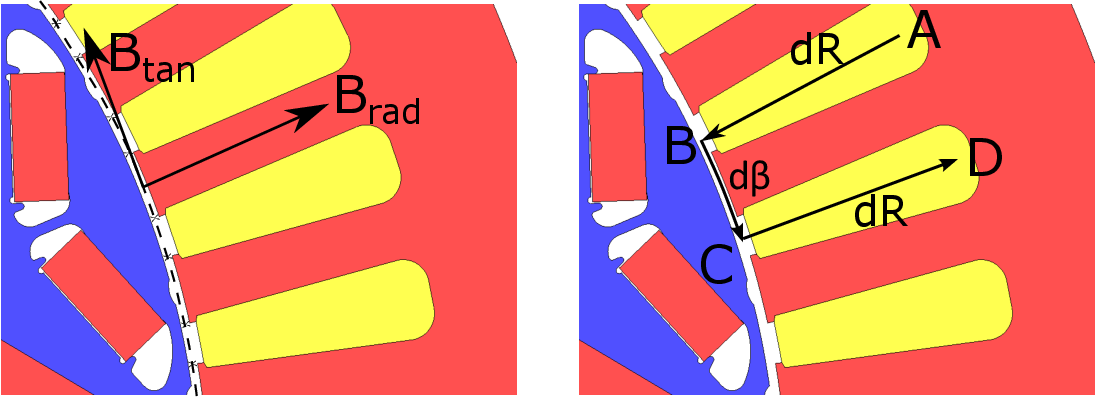}
	\end{center} 
	\caption{Calculation of radial and azimuthal force density by one section (left) or by three sections (right)}
	\label{fig:Berechnungsvarianten-MST} 
\end{figure}

\begin{figure}
	\begin{center}
		
		\includegraphics[width=0.6\textwidth]{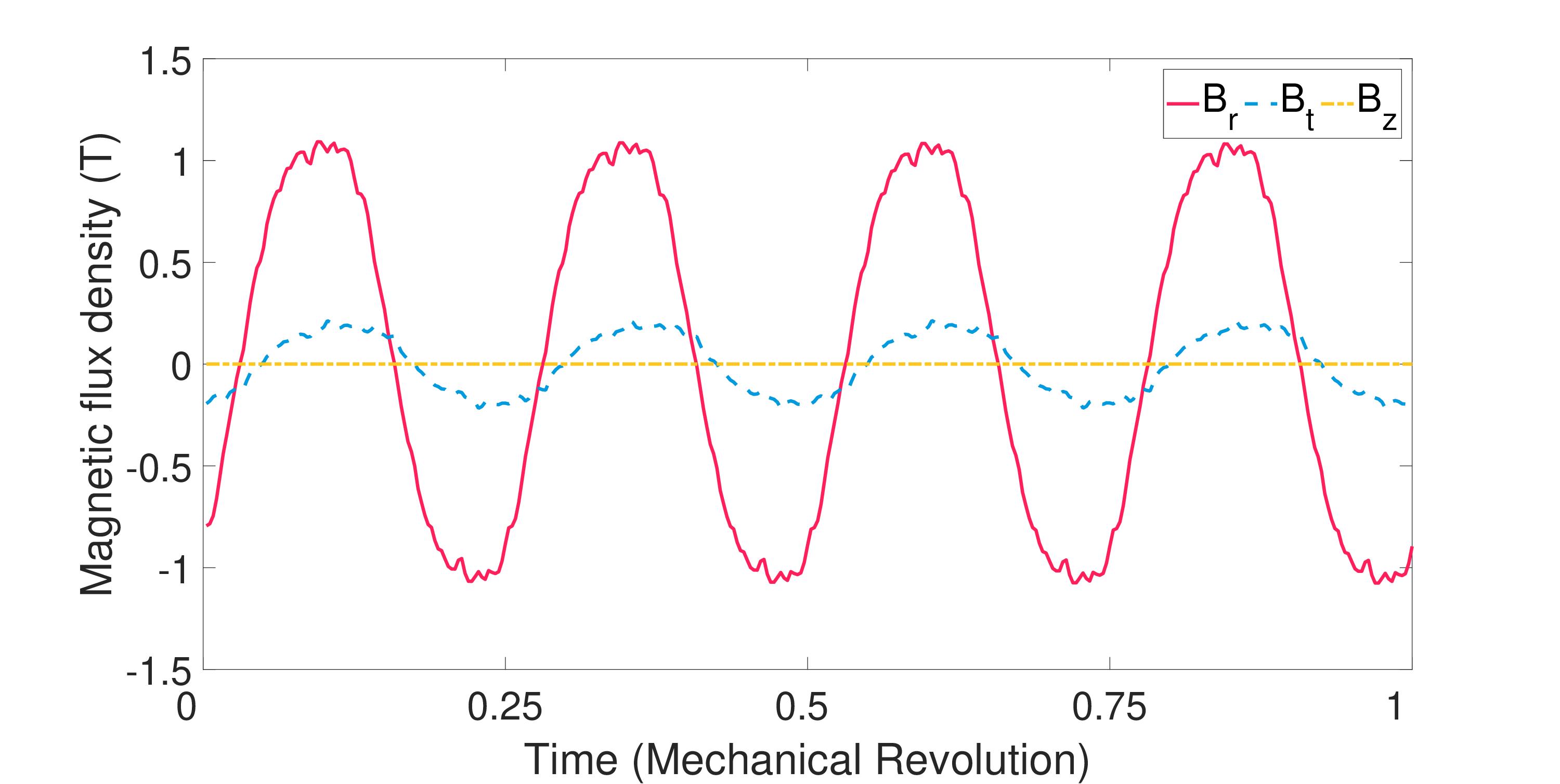}
	\end{center} 
	\caption{Magnetic flux density on a tooth for a non-skewed configuration operated at load condition.}
	\label{fig:BFeld-zeitlich} 
\end{figure}

\begin{figure}
	\centering	
	\includegraphics[width=0.95\textwidth]{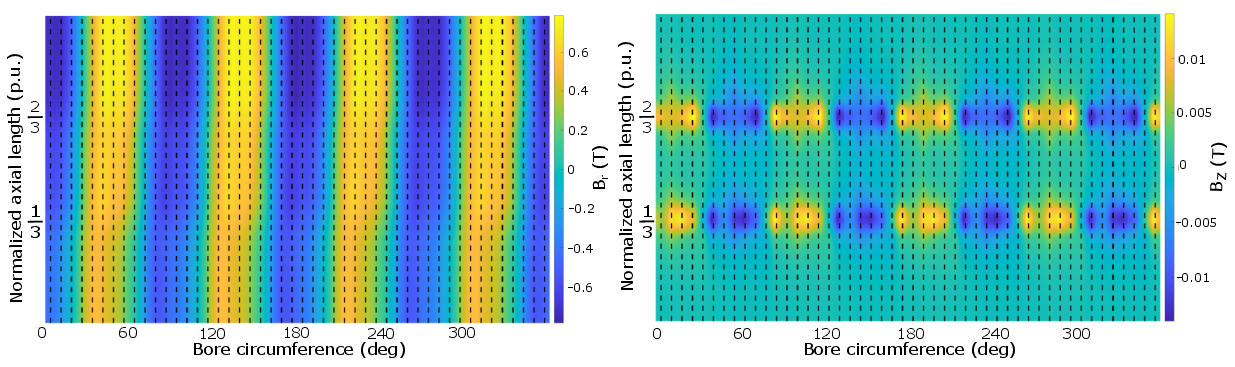}
	\caption{Radial magnetic flux density at stator teeth over axial length and bore circumference for a step-skewed rotor with three segments (left) and axial magnetic flux density (right).} 
	\label{fig:3D-Br-Bz}
\end{figure}

\begin{figure}
	\begin{center}
		
		\includegraphics[width=0.5\textwidth]{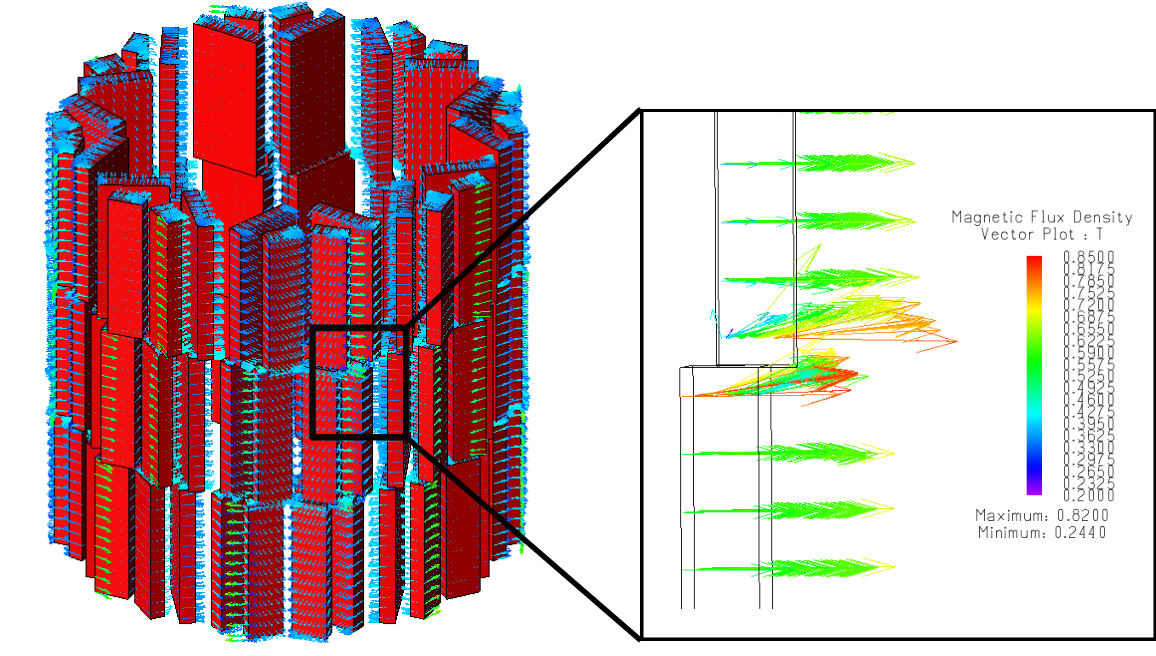}
	\end{center} 
	\caption{Direction of the magnetic flux densities at the interface between two segments of a step-skewed rotor}
	\label{fig:3D-Brichtung} 
\end{figure}

\begin{figure}
	\begin{center}
		
		\includegraphics[width=0.9\textwidth]{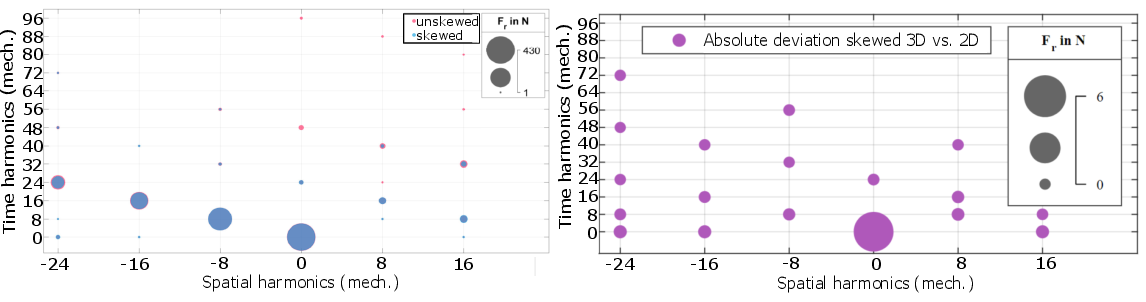}
	\end{center} 
	\caption{Mean tooth values of 2D FFT of the radial force of a step-skewed and non-skewed rotor at load condition in 2D FE (left), absolute deviation of radial tooth force between 2D and 3D FEA at load condition (right)}
	\label{fig:3D-FFT-skewed} 
\end{figure}

\begin{figure}[h]
	\begin{center}
		
		\includegraphics[width=0.8\textwidth]{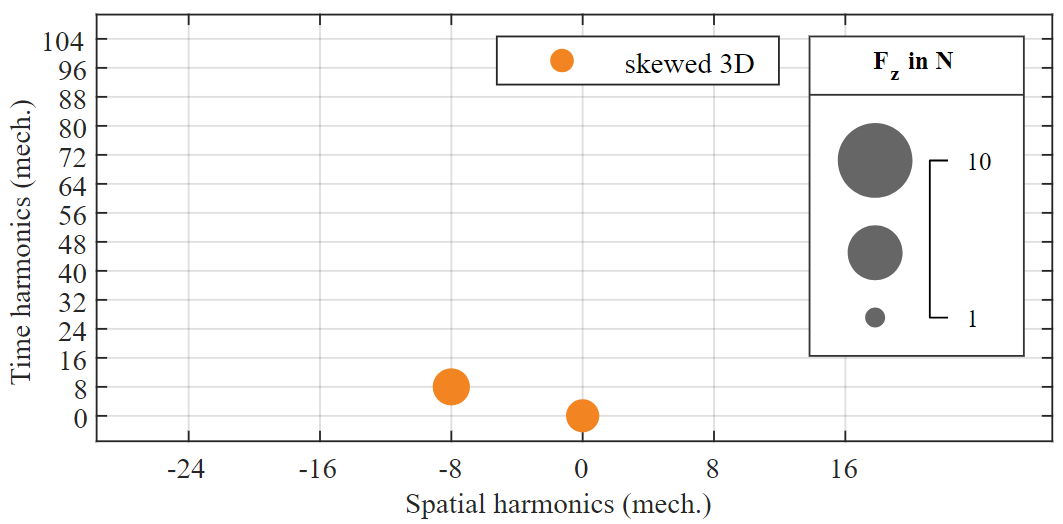}
	\end{center} 
	\caption{Mean tooth values of 2D-FFT of the radial force of a step skewed rotor at load condition in 3D FEA}
	\label{fig:3D-FFT-Fz} 
\end{figure}

The Maxwell stress tensor (MST) can be used to calculate the radial, tangential and, in 3D, also the axial force densities. Rotor skewing cause additional axial forces in RFMs that affect the NVH behaviour and bearing load of an electrical machine and are neglected in 2D FE simulations \cite{Park.2016b}. In 3D, the Maxwell stress tensor in cylindrical coordinates in the air gap is \cite{Yang.2022} 
\begin{equation}
	T_{M S}=\frac{1}{\mu_0}\left[\begin{array}{ccc}
		\frac{B_r^2-B_\theta^2-B_z^2}{2} & B_r B_\theta & B_r B_z \\
		B_\theta B_r & \frac{B_\theta^2-B_r^2-B_z^2}{2} & B_\theta B_z \\
		B_z B_r & B_z B_\theta & \frac{B_z^2-B_r^2-B_\theta^2}{2}
	\end{array}\right]
	\label{equ:Zylinderkoordinaten}
\end{equation}
where $\mu_0$ is the permeability of vacuum and $B_{\textit{r}}$, $B_{\theta}$ and $B_{\textit{z}}$ are the radial, azimuthal and axial components of the magnetic flux density. Because $B_{\mathrm{z}}$ is substantially smaller than $B_{\mathrm{r}}$ and $B_{\mathrm{t}}$, $B_{\mathrm{z}}$ is comonly neglected \cite{Kometani.1996}. The stress $\overrightarrow{\sigma}$ acting on the surface of a body $\mathcal{V}$ is
\begin{equation}
	\overrightarrow{\sigma}=T_{M S} \cdot \vec{n}
	\label{equ:sigma}
\end{equation}
with $\overrightarrow{n}$ the unit vector normal to the body's surface.
The force exerted on $\mathcal{V}$ is
\begin{equation}
	\overrightarrow{F}=\oint_{\partial\mathcal{V}} T_{M S} \cdot \vec{n} \ dS
	\label{equ:Gauss}
\end{equation}
In 2D, this simplifies into the enveloping surface $S$, which can be implemented as a circular integration line in the air gap according to Fig.~\ref{fig:Berechnungsvarianten-MST}. 
The radial, azimuthal and axial forces in 3D are then given by 
\begin{align}
	& F_{\mathrm{r}} = \frac{1}{2\mu_0} \oint_S (B_{\mathrm{r}}^2-B_{\mathrm{t}}^2-B_{\mathrm{z}}^2) r_{\delta}L \space \mathrm{d}\theta \\ 
	& F_t = \frac{1}{\mu_0} \oint_S (B_{\mathrm{t}} B_{\mathrm{r}}) r_{\delta}L\mathrm{d}\theta \\
	& F_z = \frac{1}{\mu_0} \oint_S (B_{\mathrm{z}} B_{\mathrm{r}})r_{\delta}L\mathrm{d}\theta 
\end{align}
with $L$ the axial length of the machine and $r_{\delta}$ the air gap radius.\\
In Fig.~\ref{fig:Berechnungsvarianten-MST}, different evaluation methods for calculating tooth forces are illustrated for a 2D example. The left picture shows an arc in the air gap at which the flux density $B_r$, $B_\theta$ is evaluated \cite{Boehm.2023}. 
 \\
In the right figure, occurring cross fluxes are considered by additional integration paths in the slots around a tooth. The air gap radius $r_\delta$ is necessary to calculate the path length from B to C. The radial force on a tooth corresponds to the normal stress integrated along the path BC, whereas the azimuthal force on the tooth corresponds to the normal stress integrated along the paths AB and CD. Additionally, the shear stresses evaluated at BC contribute to the azimuthal force on the tooth, while the shear stresses evaluated at AB and CD contribute to the radial force on the tooth. For 3D FE simulations, a cylindrical evaluation plane is used, on which the magnetic flux densities are decomposed into their individual components and used for force calculation.
In addition, the axial force on a skewed rotor can be calculated according to \cite{Ponick} from the torque $M$ of the machine, the rotor diameter $D$ and the skewing angle $\theta_{\mathrm{skew}}$ 
\begin{equation}
	F_z= F_t \tan \theta_{\mathrm{skew}} = \frac{2M}{D} \tan \theta_{\mathrm{skew}}
	\label{equ:Axialkraft}
\end{equation}
Various preliminary studies have shown that the deviation between the results is small due to minor magnetic cross fluxes. Therefore, the simplified approach will be pursued further due to its shorter calculation and evaluation time. \\
Figure~\ref{fig:BFeld-zeitlich} displays the magnetic flux densities over a mechanical period for a low-load operating point. For reasons of periodicity, evaluating one electrical period would be sufficient. Although the amplitude of the axial magnetic flux density is very small compared to the radial flux density, according to \cite{Ponick}, even small stresses at a corresponding frequency can lead to a perceptible NVH deterioration.
The magnetic flux densities are evaluated more precisely in Fig.~\ref{fig:3D-Br-Bz} on a cylindrical surface in the air gap for the step-skewed rotor. The stator teeth are marked by vertical black dashed lines. 
While \cite{Dupont.2014} assumes a constant radial and tangential force distribution on the rotor and stator over the axial length, Fig.~\ref{fig:3D-Br-Bz} indicates on the left side that the radial magnetic flux density distribution varies due to the inclination of the rotor by the respective skewing angles $\theta_{\mathrm{skew,q}}$. This also leads to an inhomogeneous force distribution on the stator teeth. For V-skewed rotors, this effect is amplified because of the larger number of segments. 
In contrast to the non-skewed or continuously skewed rotor, there is a local increase in the axial magnetic flux density at the transition points between individual magnetic segments (Fig.~\ref{fig:3D-Br-Bz}). For V-skewed rotors, the axial forces within the rotor cancel out each other which cause them to be to a more common choice in automotive applications. \\
Figure~\ref{fig:3D-Brichtung} shows the direction of the magnetic flux density for the step-skewed rotor in 3D FE simulation at the low-load operating point. 
The findings from Fig.~\ref{fig:3D-Br-Bz} illustrate that the axial magnetic flux density only becomes significant at the transitions between the segments. This localized effect must be considered when selecting permanent magnets to prevent demagnetization.
The spatial and temporal variations of the magnetic flux density are effectively represented as a 2D Fourier series, where the size of the dots corresponds to the magnitudes \cite{Boehm.2023} (Fig.~\ref{fig:3D-FFT-skewed}). 
Fig.~\ref{fig:3D-FFT-skewed} (left) presents the spatial harmonics as multiples of the pole number $N_p$, and the time harmonics as multiples of the stator slot number $N_s$. Data for both a non-skewed rotor and a linearly step-skewed rotor under load conditions are compared.
The dominant harmonics remain largely unaffected by the skewing. However, certain individual harmonics of the radial force, such as the 48th harmonic, are significantly reduced, similar to the reduction observed for the torque harmonics.
The absolute deviation between the 2D and 3D simulation results for the harmonics is shown in Fig.~\ref{fig:3D-FFT-skewed} (right). With a maximum radial force of $430 \unit{N}$, the observed deviation is $10 \unit{N}$, which corresponds to a discrepancy of less than $<3 \%$.
The axial force resulting from the axial field components (Fig.~\ref{fig:3D-Brichtung}) is shown in Fig.~\ref{fig:3D-FFT-Fz}. This additional force can only be obtained from 3D FE simulations.

\section{Conclusion}
Different skewing configurations of a PMSM are evaluated. Rotor skewing reduces field disturbances and the amplitudes of torque harmonics, which can be observed in both 2D and 3D simulations, with the 2D FE results being sufficient for most analyses. At the transition points between individual magnetic segments, axial field components emerge, leading to additional force contributions in the axial direction. These axial forces are only detectable in 3D simulations. For NVH (Noise, Vibration, and Harshness) analysis, the radial forces are particularly evaluated. The forces calculated using the Maxwell stress tensor serve as input data for structural dynamic simulations.
Meaningful results can be obtained from 2D FE simulations. However, axial force components, relevant for bearing design and demagnetization effects, require the application of 3D FE simulations.

\bibliographystyle{./IEEEtran}
\bibliography{authorref}



\end{document}